\documentclass[%twocolumn,
secnumarabic, amssymb, amsmath, nofootinbib,tightenlines, showpacs, nobibnotes, aps, prl]{revtex4}
\usepackage{amsfonts}
\usepackage{docs}%
\usepackage{bm}%
\usepackage[colorlinks=true,linkcolor=blue]{hyperref}%
\usepackage{graphicx}% Include figure files
\usepackage{dcolumn}% Align table columns on decimal point

\begin{document}

\title{Orbital Diamagnetism of Weak-doped Bilayer Graphene in Magnetic Field}

\author{Min Lv}
\author{Shaolong Wan}
\altaffiliation{Corresponding author} \email{slwan@ustc.edu.cn}
\affiliation{Institute for Theoretical Physics and Department of Modern Physics \\
University of Science and Technology of China, Hefei, 230026, {\bf
P. R. China}}

\date{\today}

\begin{abstract}
We investigate the orbital diamagnetism of a weak-doped bilayer
graphene (BLG) in spatially smoothly varying magnetic field and
obtain the general analytic expression of the orbital
susceptibility of BLG, with finite wave number and Fermi energy,
at zero temperature. We find that the magnetic field screening
factor of BLG is dependent with the wave number, which results in
a more complicated screening behavior compared with that of
monolayer graphene (MLG). We also study the induced magnetization,
electric current in BLG, under nonuniform magnetic field, and find
that they are qualitatively different from that in MLG and
two-dimensional electron gas (2DEG). However, similar to the MLG,
the magnetic object placed above BLG is repelled by a diamagnetic
force from BLG, approximately equivalent to a force produced by
its mirror image on the other side of BLG with a reduced amplitude
dependent with the typical length of the systems. BLG shows
crossover behaviors in the responses to the external magnetic
field as the intermediate between MLG and 2DEG.
\end{abstract}


\maketitle

\section{Introduction}

Bilayer graphene (BLG), as a significant graphene-related
material, has attracted much attentions \cite{McCann, Morozov,
Novoselov-2} due to its unusual electronic structure. Formed by
stacking two monolayer graphene (MLG) in Bernal stacking, Bilayer
graphene has four inequivalent sites in each unit cell, including
$A_1$ and $B_1$ atoms on the top layer and $A_2$ and $B_2$ atoms
on the bottom layer. The distance $a$ between $A_1$ and $B_1$ is
about $0.142$ nm and the vertical separation of the two layers $d$
is about $0.334$ nm. On its band structure side, around the point
where the conduction and valence band touch, BLG has a quadratic
energy dispersion \cite{McCann} similar to the regular
two-dimensional electron gas (2DEG) but its low-energy effective
Hamiltonian is chiral without bandgap similar to the MLG. Another
unique feature of BLG is that a widely tunable bandgap can realize
conveniently by introducing an electrostatic potential bias
between the up and bottom layer \cite{McCann, Min, McCann-2, Lu,
Guinea, Zhang}.

The magnetic susceptibility of electronic systems comes from two
contribution: One is Pauli paramagnetism which stems from spin
polarization; The other is Landau diamagnetism which stems from
the circulation of orbital currents. The orbital diamagnetism of
carbon systems has attracted the interest of both experimental and
theoretical physicists for a long time. It was firstly found by
Krishnan \cite{Krishnan} that the diamagnetic susceptibility of
the bulk graphite is large and anisotropic. McClure \cite{McClure}
showed it arises from the Landau quantization of 2D massless Dirac
fermions, which results in a delta function peak at zero energy in
the orbital diamagnetism of graphite. After the experimentally
fabrication of graphene, a great deal of works have concerned
about the orbital magnetism of graphene-related systems, such as
nodal fermions \cite{Ghosal}, disordered graphene \cite{Sharapov,
Fukuyama, Koshino-1}, few-layered graphene \cite{Koshino-2,
Nakamura, Neto}, graphene in nonuniform magnetic field
\cite{Koshino-3, Principi}. Quite recently, novel paramagnetic
susceptibility has been found in MLG and BLG with bandgap
\cite{Koshino-4} and doped graphene to the first order in the
Coulomb interaction \cite{Principi-2}.

Safran \cite{Safran}, Koshino and Ando \cite{Koshino-2} have
derived the analytical expression of orbital susceptibility
$\chi(0)$ for bilayer graphene, but their results are just limited
to the case of zero wave number. In this article we generalize
them to general cases and analytically study the orbital
diamagnetism of a weak-doped BLG (i.e., the Fermi energy
$\epsilon_F$ is much smaller than the interlayer hopping energy.)
in nonuniform magnetic fields.

This article is organized as follows. In Sec. II, the effective
Hamiltonian of bilayer graphene and its corresponding eigenstates
and eigenenergies are introduced. In Sec. III, the orbital
susceptibility of bilayer graphene is studied. In Sec. IV, we
investigate and discuss the responses of bilayer graphene to
several specific external magnetic field. The conclusion is given
in Sec. V.

\section{Bilayer Graphene Effective Model }

In the low energy and long wave regime, the BLG Hamiltonian near a
$K$ point, in the absence of a magnetic field, can written as an
excellent approximate form \cite{McCann} (we set $c = 1 = \hbar$
in this paper):
\begin{equation}
H_0 =  \left(
\begin{array}{cccc}
0 & \gamma \hat{k}_{-} & 0 & 0 \\
\\
\gamma \hat{k}_{+} & 0 & \Delta & 0 \\
\\
0 & \Delta & 0 & \gamma \hat{k}_{-} \\
\\
0 & 0 & \gamma \hat{k}_{+} & 0
\end{array} \right), \label{2.1}
\end{equation}
where $\gamma=3ta/2\approx 10^6 m/s$ is the monolayer graphene
Fermi velocity, $t\approx 3eV$ is the in-plane hopping energy,
$a\approx 0.142$ nm is the in-plane interatomic distance,
$\Delta\approx 0.35eV$ is the interlayer hopping energy.
$\hat{\bm{k}}=(\hat{k}_x,\hat{k}_y)=-i\bm{\nabla}$ is a 2D
wave-vector operator, $\hat{k}_{\pm}=\hat{k}_x\pm i\hat{k}_y$. In
order to solve the eigen equation of the Hamiltonian (\ref{2.1}),
the wave function can be expressed as $(\psi_{A_1},\psi_{B_1},
\psi_{A_2},\psi_{B_2})$, where the four components represent the
Bloch functions at $A_1$, $B_1$, $A_2$ and $B_2$ sites,
respectively. Follow the stipulation of Ando \cite{Ando}, and
define
\begin{eqnarray}
\epsilon(k) &=& \sqrt{(\frac{\Delta}{2})^2+(\gamma k)^2},\\
\gamma k &=& \epsilon(k)\sin{\psi},\\
\frac{\Delta}{2} &=& \epsilon(k)\cos{\psi}. \label{2.2}
\end{eqnarray}
Then the corresponding eigenstates of Eq.(\ref{2.1}) are given as
\begin{eqnarray}
\Psi_{sj{\bm{k}}} (\bm{r}) = \frac{1}{L}\exp(i \bm{k} \cdot
\bm{r})U[\theta_{\bm{k}}] F_{sjk}, \label{2.3}
\end{eqnarray}
where $L^2$ is the area of the system, $s=+1$ and $-1$ denote the
conduction and valence bands, respectively, $j=1$ and $2$
specifies two subbands within the conduction or valence bands,
$\theta_{\bm{k}}=\arctan (k_y/k_x)$ is the polar angle of the
momentum $\bm{k}$,
\begin{equation}
U(\theta) =  \left(
\begin{array}{cccc}
1 & 0 & 0 & 0 \\
\\
0 & e^{i\theta} & 0 & 0 \\
\\
0 & 0 & e^{i\theta} & 0 \\
\\
0 & 0 & 0 & e^{2i\theta}
\end{array} \right), \label{2.4}
\end{equation}
and
\begin{eqnarray}
F_{s1k} =\frac{1}{\sqrt{2}} \left(
\begin{array}{c}
s\cos(\psi/2) \\
\\
\sin(\psi/2)\\
\\
-s\sin(\psi/2)\\
\\
-\cos(\psi/2)
\end{array} \right), \label{2.5a}
~~~~~~~~~~F_{s2k} =\frac{1}{\sqrt{2}} \left(
\begin{array}{c}
s\sin(\psi/2) \\
\\
\cos(\psi/2)\\
\\
s\cos(\psi/2)\\
\\
\sin(\psi/2)
\end{array} \right).
\label{2.5b}
\end{eqnarray}
The corresponding eigenenergies of Eq.(\ref{2.1}) are
\begin{eqnarray}
\epsilon_{s1}(k) &=& 2s\epsilon(k)\sin^2(\psi/2), \nonumber\\
\epsilon_{s2}(k) &=& 2s\epsilon(k)\cos^2(\psi/2). \label{2.6}
\end{eqnarray}

Considering a magnetic field $\bm{B}(\bm{r})=[\nabla \times
\bm{A}(\bm{r})]_z$, the Hamiltonian for the system is: $H = H_0 +
H_1$, with $H_1 = - \int d^2 r j_{\alpha}(\bm{r})
A_{\alpha}(\bm{r}, t)$. The current operator at $\bm{r}_0$ is
given
\begin{eqnarray}
\hat{j}_{\alpha}(\bm{r_0})=\frac{e}{2}[\hat{v}_{\alpha}\delta(\bm{r}-\bm{r}_0)+
\delta(\bm{r}-\bm{r}_0)\hat{v}_{\alpha}],~~~~~~~~~~~~~~~~(\alpha =
x, y) \label{2.7}
\end{eqnarray}
where $\hat{v}_{\alpha}$ is velocity operator
\begin{eqnarray}
\hat{v}_{\alpha}=\frac{\partial H_0}{\partial k_{\alpha}}=
\gamma\left(
\begin{array}{cc}
\sigma_{\alpha} & 0 \\
\\
0 & \sigma_{\alpha}
\end{array} \right), ~~~~~~~~~~~~~~~~(\alpha = x, y)
\label{2.8}
\end{eqnarray}
$\sigma_{x,y}$ are the Pauli matrices which act on the sublattice
space within a layer.

\section{Orbital Susceptibility}

The finite wave number susceptibility $\chi(\bm{q})$ can be
obtained through the Kubo formula \cite{Mahan}. Within the linear
response theory, the external vector potential $\bm{A}$ and the
its induced 2D electric current density $\bm{j}$ have a relation
\begin{eqnarray}
j_{\mu}(\bm{q})=\sum_{\nu}K_{\mu\nu}(\bm{q})A_{\nu}(\bm{q}),
\label{3.9}
\end{eqnarray}
and the orbital susceptibility $\chi(\bm{q})$ and the response
tensor $K_{\mu\nu}(\bm{q})$ are related by
\begin{eqnarray}
K_{\mu\nu}(\bm{q})=q^2\chi(\bm{q})(\delta_{\mu\nu}-\frac{q_{\mu}q_{\nu}}{q^2}).
\label{3.10}
\end{eqnarray}

In the first order perturbation, we have
\begin{eqnarray}
K_{\mu\nu}(\bm{q})=-\frac{g}{L^2}\sum_{ss'jj'\bm{k}}
\frac{f[\epsilon_{sj}(k)]-f[\epsilon_{s'j'}(k')]}{\epsilon_{sj}(k)-\epsilon_{s'j'}(k')}I_{ss',jj'},
\label{3.11}
\end{eqnarray}
where $g=g_vg_s=4$ is the total degeneracy,
$\bm{k'}=\bm{k}+\bm{q}$, $\epsilon_{sj}(k)$ is the eigenenergy
given by Eq.(\ref{2.6}), $f(\epsilon)$ is the Fermi distribution
function $f(\epsilon) = [1 + \exp{\beta(\epsilon -
\epsilon_F)}]^{-1}$ where $\epsilon_F$ is the Fermi energy,
$\beta=1/(k_BT)$. $I_{ss',jj'}$ is the current-current response
matrix element expressed by
\begin{eqnarray}
I_{ss',jj'}=\left[F^{\dagger}_{s'j'\bm{k'}}U^{\dagger}(\theta_{\bm{k'}})v_{\mu}U(\theta_{\bm{k}})F_{sj\bm{k}}\right]
\left[F^{\dagger}_{sj\bm{k}}U^{\dagger}(\theta_{\bm{k}})v_{\nu}U(\theta_{\bm{k'}})F_{s'j'\bm{k'}}\right],
\label{3.12}
\end{eqnarray}
which determines the weight of contribution of the transition from
subband $j$ to $j'$, with $ss' = +1$ and $-1$ denote the intraband
and interband transition, respectively.

Define the effective mass $m\equiv \Delta/(2\gamma^2)\approx
0.033m_e$ and the Fermi wave number $k_F\equiv
\sqrt{2m\epsilon_F}$. For a low-energy theory of bilayer graphene,
there is a natural high-energy cutoff wave number $\Lambda \equiv
\sqrt{2m\Delta} = \Delta/{\gamma}$. When $k_F,q \ll \Lambda$,
i.e., when bilayer graphene is weak-doped and the external field
is smooth enough compared to the cutoff wavelength, the
transitions intra the same $j$ subband dominate the contribution
to the response function. One of these transitions is the
transition intra the $j=1$ subband:

\begin{eqnarray}
K_{\mu\nu,11}(\bm{q})=-\frac{g}{L^2}\sum_{ss'\bm{k}}
\frac{f[\epsilon_{s1}(k)]
-f[\epsilon_{s'1}(k')]}{\epsilon_{s1}(k)-\epsilon_{s'1}(k')}I_{ss',11}.
\label{3.13a}
\end{eqnarray}
In the low energy limit, it can be approximately given as

\begin{eqnarray}
K_{\mu\nu,11}(\bm{q})\approx-\frac{ge^2}{m^2L^2}\sum_{ss'\bm{k}}\frac{f[s\epsilon_k]-f[s'\epsilon_{k'}]}{s\epsilon_k-s'\epsilon_{k'}}
\left[\frac{1}{2}(\bm{k}+\frac{\bm{q}}{2})^2\delta_{\mu\nu}+\frac{ss'}{8}F_{\mu\nu}\right],
\label{3.13b}
\end{eqnarray}
with $\epsilon_k=k^2/(2m)$ and
\begin{eqnarray}
F_{\mu\nu}=(\delta_{\mu1}\delta_{\nu1}-\delta_{\mu2}\delta_{\nu2})\left[k^2\cos{2\theta_{\bm{k'}}}+k'^2\cos{2\theta_{\bm{k}}}
+2kk'\cos(\theta_{\bm{k}}+\theta_{\bm{k'}})\right]\nonumber
\\+(\delta_{\mu1}\delta_{\nu2}+\delta_{\mu2}\delta_{\nu1})
\left[k^2\sin{2\theta_{\bm{k'}}}+k'^2\sin{2\theta_{\bm{k}}}+2kk'\sin(\theta_{\bm{k}}+\theta_{\bm{k'}})\right].
\label{3.14}
\end{eqnarray}
The other is the transition intra the subband $j=2$:

\begin{eqnarray}
K_{\mu\nu,22}(\bm{q})=-\frac{g}{L^2}\sum_{ss'\bm{k}}
\frac{f[\epsilon_{s2}(k)]
-f[\epsilon_{s'2}(k')]}{\epsilon_{s2}(k)-\epsilon_{s'2}(k')}I_{ss',22}.
\label{3.15a}
\end{eqnarray}
By using the fact that the subband $j=2$ in the conduction band
($s = + 1$) is empty in the weak-doped limit, it can be
approximately given as

\begin{eqnarray}
K_{\mu\nu,22}(\bm{q})\approx
\frac{2ge^2}{m^2L^2}\sum_{\bm{k}}\frac{1}{\epsilon_k-\epsilon_{k'}}
\left[\frac{1}{2}(\bm{k}+\frac{\bm{q}}{2})^2\delta_{\mu\nu}+\frac{ss'}{8}G_{\mu\nu}\right],
\label{3.15b}
\end{eqnarray}
with
\begin{eqnarray}
G_{\mu\nu}=(\delta_{\mu1}\delta_{\nu1}-\delta_{\mu2}\delta_{\nu2})\left[k^2\cos{2\theta_{\bm{k}}}+k'^2\cos{2\theta_{\bm{k'}}}
+2kk'\cos(\theta_{\bm{k}}+\theta_{\bm{k'}})\right]\nonumber
\\+(\delta_{\mu1}\delta_{\nu2}+\delta_{\mu2}\delta_{\nu1})
\left[k^2\sin{2\theta_{\bm{k}}}+k'^2\sin{2\theta_{\bm{k'}}}+2kk'\sin(\theta_{\bm{k}}+\theta_{\bm{k'}})\right].
\label{3.16}
\end{eqnarray}

From the above, we firstly obtain the analytic expression of the
susceptibility of bilayer graphene at zero temperature as:
\begin{eqnarray}
\chi(\bm{q};\epsilon_F)=\frac{ge^2}{8\pi
m}\left\{\log{\frac{2k^2_F+\sqrt{4k^4_F+q^4}}{4\Lambda^2}}+\frac{1}{3}\left[1+
(1-\frac{4k^2_F}{q^2})^{3/2}\theta(q-2k_F) \right]\right\},
\label{3.17}
\end{eqnarray}
where $\theta(x)$ is the step function defined by $\theta(x)=1
(x>0)$ and $0 (x<0)$, which is a centra result of our work. From
Eq.(\ref{3.17}), on the one hand, we can give the orbital
susceptibility of BLG at zero wave number as
\begin{eqnarray}
\chi(\bm{q}=0;\epsilon_F)=\frac{ge^2}{8\pi
m}\left[\log{\frac{\epsilon_F}{\Delta}}+\frac{1}{3} \right].
\label{3.18}
\end{eqnarray}
This result is same as the result given in \cite{Safran,
Koshino-2}, which shows a logarithmically diverging behavior at
$\epsilon_F = 0$. On the other hand, we can give the orbital
susceptibility of BLG at zero Fermi energy as
\begin{eqnarray}
\chi(\bm{q};\epsilon_F=0)=\frac{ge^2}{4\pi
m}\left[\log{\frac{q}{2\Lambda}}+\frac{1}{3} \right]. \label{3.19}
\end{eqnarray}
This also shows a logarithmically diverging behavior at $q=0$. It
is easy to see that just by replacing $\epsilon_F$ with $\gamma
q/2$ and increasing the coefficient to its double times, we can
transform the susceptibility $\chi(\bm{q} = 0; \epsilon_F)$ into
$\chi(\bm{q}; \epsilon_F = 0)$.

The orbital magnetic susceptibility $\chi(\bm{q})$ of bilayer
graphene as a function of wave number is shown in Fig.[1]. For
comparing the wave number-dependent behaviors of susceptibility of
the MLG, BLG and 2DEG systems, we provide below the finite wave
number susceptibility of MLG and 2DEG, which has been given in
Ref.\cite{Koshino-3},
\begin{eqnarray}
\chi(\bm{q};\epsilon_F)=-\frac{ge^2v}{16}\frac{1}{q}\theta(q-2k_F)\left[1+\frac{2}{\pi}
\frac{2k_F}{q}\sqrt{1-\left(\frac{2k_F}{q}\right)^2}-\frac{2}{\pi}\sin^{-1}\frac{2k_F}{q}
\right]~~~~~~~~~~~(for~MLG ), \label{3.20}
\end{eqnarray}
\begin{eqnarray}
\chi(\bm{q};\epsilon_F)=\frac{ge^2}{24\pi
m}\left[\left(1-\frac{4k^2_F}{q^2}\right)^{3/2}\theta(q-2k_F)
-1\right]~~~~~~~~~~~(for~2DEG ). \label{3.21}
\end{eqnarray}

At $q=0$, the susceptibility of BLG $-\chi(0,\epsilon_F)\propto
\log(\Delta/\epsilon_F)$ is rather different from that of MLG,
where $-\chi(0, \epsilon_F) \propto \delta(\epsilon_F)$, and 2DEG,
where $-\chi(0, \epsilon_F) \propto 1$. By comparing their
diverging behaviors, it can be found that the BLG in some sense
shows an intermediate behavior between the MLG and 2DEG. For small
$q$, the $-\chi(\bm{q};\epsilon_F)$ of BLG deviates from the
$-\chi(0;\epsilon_F)$ as $(q/2k_F)^4$, and falls more rapidly as
$q$ increase. On the other hand, the susceptibility of MLG
vanishes while 2DEG maintains as a constant for the whole regime
$q<2k_F$ (see Fig.[1] of Ref. \cite{Koshino-3}). At $q=2k_F$, the susceptibility $\chi(2k_F;\epsilon_F)$
of MLG and 2DEG are both constants (zero for MLG) which are
independent of the Fermi wave number $k_F$, but for BLG, we have
\begin{eqnarray}
\chi(2k_F;\epsilon_F)=\frac{ge^2}{8\pi
m}\left(\log{\frac{\epsilon_F}{\Delta}}+\frac{1}{3}+\log{\frac{1+\sqrt{5}}{2}}\right),
\label{3.22}
\end{eqnarray}
which is dependent of the $k_F$. In contrast to the MLG, the
susceptibility of BLG has no singular behavior at $q=2k_F$, and it
is continuous as well as its first derivative. For large $q$,
especially for $q\gg 2k_F$, $-\chi(q)$ of BLG rapidly approaches the
curve of Eq. (\ref{3.19}) and falls as
$\log(1/q)$, very different from that of MLG where the
susceptibility falls off more rapidly ($\sim 1/q$) and 2DEG where
it falls as $1/q^2$. Due to having the same parabolic energy
dispersion in the low energy limit, susceptibility of Bilayer
graphene and 2DEG share the same term
$ge^2(1-4k^2_F/q^2)^{3/2}\theta(q-2k_F)/(24\pi m)$.

\begin{figure}
\includegraphics[width=8.5cm, height=7cm]{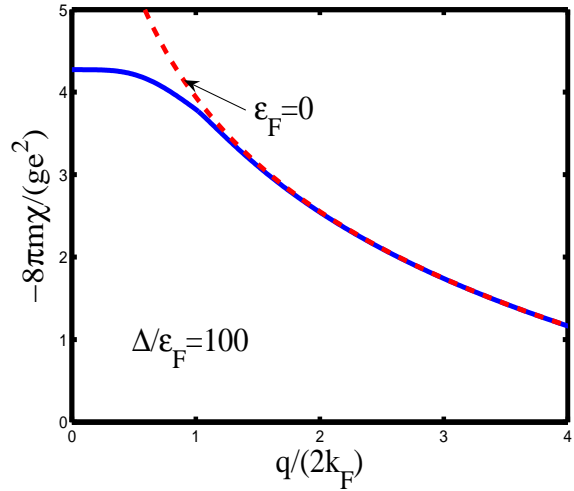}
\caption{Magnetic susceptibility $\chi(\bm{q})$ in bilayer
graphene. Here we use $\Delta/\epsilon_F=100$.} \label{fig.1}
\end{figure}

\section{Responses To Specific External Magnetic Field}

Now we study the responses of bilayer graphene to different types
of magnetic field. First let us consider the case of a neutral BLG
(i.e.,$~\epsilon_F=0$) under a sinusoidal magnetic field
$\bm{B}(\bm{r})=B_0\cos{qx}\bm{e}_z$. Defining
$\chi(q)\equiv\chi(\bm{q},\epsilon_F=0)$, we have the induced
magnetization $\bm{m}(\bm{r})=\chi(q)\bm{B}(\bm{r})$, the induced
current $\bm{j}(\bm{r})=q\chi(q)B_0\sin{qx}\bm{e}_y$, and the $z$
component of induced counter magnetic field on BLG
\begin{eqnarray}
\bm{B}_{ind}(\bm{r})=-\alpha_g(q)\bm{B}(\bm{r}),\label{4.23}
\end{eqnarray}
with the magnetic field screening factor
\begin{eqnarray}
\alpha_g(q)=-\frac{ge^2q}{2m}\left[\log{\frac{q}{2\Lambda}}+\frac{1}{3}\right].
\label{4.24}
\end{eqnarray}
When $q\rightarrow 0$, we have $\alpha_g(q)\rightarrow 0$; i.e.,
under a constant magnetic field there is no counter magnetic field
on BLG, which is different from that of MLG. In MLG, the magnetic
field screening factor is fixed and independent of $q$ and the
specific form of external field, the induced magnetic field above
the graphene layer is simply equivalent to the field of a mirror
image of the original object reflected with respect to the
graphene layer but reduced by $\alpha_g$ \cite{Koshino-3}.
However, this argument does not fit for the case of BLG, since the
magnetic field screening factor of BLG is dependent of $q$ and
complicated.

Next we consider the case of a line current $I$ flowing along the
$+ y$ direction above the BLG, and passing through the point $(0,
0, d)~(d > 0)$. The z component of the magnetic field on BLG is
$B(\bm{r}) = - 2I x/(x^2 + d^2)$. For $\epsilon_F = 0$, the
induced magnetization can be given as
\begin{eqnarray}
m(\bm{r})=-\frac{Ige^2}{2\pi
m}\frac{x}{x^2+d^2}\left[\log(2\Lambda\sqrt{x^2+d^2})+
\frac{d}{x}\arctan{\frac{x}{d}}-\gamma_E+\frac{1}{3}\right],
\label{4.25}
\end{eqnarray}
here $\gamma_E\approx 0.577$ is the Euler constant. By using
$\bm{j}_{ind}=\nabla\times \bm{m}(\bm{r})$, we obtain the induced
electric current $j_{ind}(\bm{r})=j_y\bm{e}_y$, where
\begin{eqnarray}
j_y=\frac{Ige^2}{2\pi
m}\frac{1}{(x^2+d^2)^2}\left\{(d^2-x^2)\left[\log(2\Lambda\sqrt{x^2+d^2})-\gamma_E+\frac{1}{3}\right]
-2dx\arctan{\frac{x}{d}}+(x^2+d^2)\right\}. \label{4.26}
\end{eqnarray}
The integral of $j_y$ in $x$ exactly equals to $0$, which means
the external electric current $I$ can not induce an effective
transport electric current on the BLG, in contrast to that of MLG,
where $I$ induces an effective electric current $-\alpha_g I$. The
induced magnetic field on BLG can be given as
$\bm{B}_{ind}(\bm{r})=B_z\bm{e}_z$ with
\begin{eqnarray}
B_z=\frac{Ige^2}{m}\frac{1}{(x^2+d^2)^2}\left[2dx\left(\log{\frac{\sqrt{x^2+d^2}}{8\Lambda
d^2}} +
\gamma_E-\frac{1}{3}\right)+(d^2-x^2)\arctan{\frac{x}{d}}\right].
\label{4.27}
\end{eqnarray}
In MLG, the induced magnetic field $- \alpha_g I x/(x^2 + d^2)$ is
equivalent to the field created by a current $- \alpha_g I$
flowing at $z = - d$. This argument does not fit for the case of
BLG as shown in Eq.(\ref{4.27}). At large distance $x\gg d$, the
induced magnetic field is proportion to $\sim 1/x^2$ comparing
with $\sim 1/x$ in MLG. However, the original current is repelled
by a force $\approx \alpha_g(1/2d)I^2/d$ per unit length, which
can be approximately but not exactly considered as a force created
by a current $\alpha_g(q) I$ at $z = - d$ with $q = 1/2d$.

As another typical example, we study the the induced
magnetization, electric current and magnetic field by a magnetic
monopole $q_m$ laying above the BLG. Suppose $q_m$ is located at
the point $(0, 0, d)$, $(d > 0)$, and the BLG plane is $z = 0$.
The magnetic field perpendicular to the BLG is given as
$B(\bm{r})=q_md/(r^2+d^2)^{3/2}$ with $r=\sqrt{x^2+y^2}$. For
neutral BLG, the induced magnetization is given by
\begin{eqnarray}
m(\bm{r})=-\frac{q_mge^2}{4\pi
md^2}\left\{F(\frac{r}{d})-(\log{2\Lambda
r}-1/3)\left[1+(\frac{r}{d})^2\right]^{-3/2}\right\}, \label{4.28}
\end{eqnarray}
where the function
\begin{eqnarray}
F(x)=\frac{1}{x^2}\int_0^{\infty}zJ_0(z)\log{z}e^{-z/x}dz .
\label{4.29}
\end{eqnarray}
At small distance $(r\ll d)$ and large distance $(r\gg d)$, the
induced magnetization can be written as
\begin{eqnarray}
m(r) = \frac{q_mge^2}{4\pi m}\times\left\{
\begin{array}{ll} \left\{\log{2\Lambda d}+\gamma_E-\frac{4}{3}-\frac{3}{2}(r/d)^2
\left[\log{2\Lambda r}-\frac{1}{3}\right]\right\}/d^2 &
( for \hspace{3mm}r\ll d) \\
\\
1/r^2 & ( for \hspace{3mm} r\gg d )
\end{array} \right. \\
\label{4.30}
\end{eqnarray}
It is interesting to see that at large distance the induced
magnetization of BLG $m(r)\propto 1/r^2$, while that of MLG
$\propto 1/r$ and 2DEG $\propto 1/r^3$; i.e., the BLG shows a
behavior as the crossover from MLG to 2DEG. The integral of $m(r)$
over the plane has a logarithmically diverging behavior
$\propto\log{R}$ when the distance $R\rightarrow \infty$.
According to Eq.(\ref{4.28}), the corresponding electric current
can be given by $\bm{j}(\bm{r})=-(\partial m/\partial
r)\bm{e}_{\theta}\equiv j_{\theta}\bm{e}_{\theta}$. When $r$ is
small or large, we obtain
\begin{eqnarray}
j_{\theta} = \frac{q_mge^2}{4\pi m}\times\left\{
\begin{array}{ll} 3r\log{2\Lambda r}/d^4 &
( for \hspace{3mm}r\ll d) \\
\\
2/r^3 & ( for \hspace{3mm} r\gg d )
\end{array} \right. \\
\label{4.31}
\end{eqnarray}
Remind that our results are confined to the limit $\Lambda r\gg
1$, and therefore $j_r$ is positive through out the realistic
distance. The current $j_{\theta}$ in MLG $\propto
r/(r^2+d^2)^{3/2}$, whose asymptotic form is $\propto r$ at small
distance and $\propto 1/r^2$ at large distance. We can find that
the induced currents $j_{\theta}$ of MLG and BLG are qualitatively
different in all distance. However, similar to the case of line
current, the force between the monopole and the BLG can be
approximately written as $\alpha_g(q) q_m^2/(2d)^2$ with $q =
1/d$, which has the same form as that of MLG $\alpha_g
q_m^2/(2d)^2$.

\begin{figure}
\includegraphics[width=8.5cm, height=7cm]{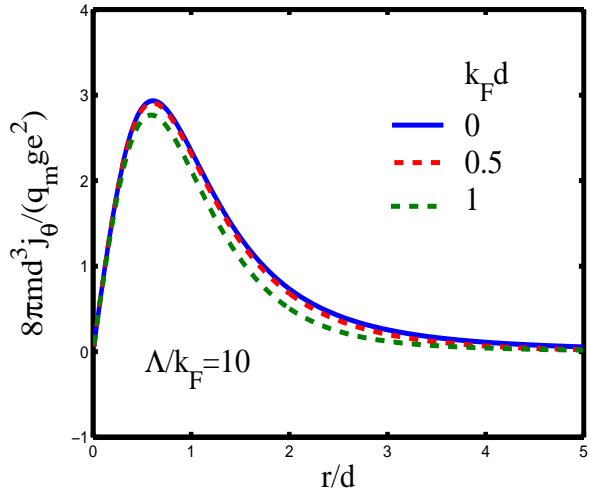}
\caption{Electric current $j_{\theta}(r)$ on bilayer graphene
induced by a magnetic charge $q_m$ at $z=d$. Here we use
$\Lambda/k_F=10$.} \label{fig.2}
\end{figure}

For doped bilayer graphene $(\epsilon_F\neq 0)$, the corresponding
response current $j_{\theta}(r)$ for different values of $k_F$ are
shown in Fig.[2]. We can find that the current change slightly as
the Fermi wave number increase from $k_F d = 0$ to $k_F d = 1$;
i.e., the induced current has a weak Fermi wave number dependence,
which is rather different from that of MLG. This arises from the
fact that the dominant term involving $k_F$ in the susceptibility
is logarithmically.

Different from the traditional 2DEG, MLG has a peculiar property
\cite{Koshino-3}: (1) The counter field induced by the response
current mimics a mirror image of the original object. (2) The
object is repelled by a diamagnetic force from the MLG, as if
there exists its mirror image with a reduced amplitude on the
other side of MLG. With the investigation above, we find that
argument (1) can be not extended to BLG. However, the argument (2)
still be approximately correct for weak-doped BLG, and it only
needs to replace the constant reduced amplitude with a reduced
amplitude dependent with the typical length of the systems. The
BLG, in some sense, still show an intermediate behavior between
the MLG and 2DEG.

It is significant to compare the contribution of Landau
diamagnetism to the whole magnetism with that of Pauli
paramagnetism in BLG. At $\epsilon_F = 0$, the Pauli spin
susceptibility $\chi^{spin}(q)$, which is given by the
density-density response function \cite{Hwang}, is equivalent to
$g_vm\mu_B^2\log4/2\pi$, where $\mu_B$ is the Bohr magneton. With
Eq.(\ref{3.19}), we obtain the ratio $\chi^{spin}/\chi^{orb}\sim
(0.03)^2/(\log{2\Lambda/q})$, which is rather small in our theory
$(for~q \ll \Lambda)$.

The temperature also has an influence on the BLG diamagnetism. For
$q=0$, this has been discussed by Safran \cite{Safran}, who shows
the susceptibility with finite temperature and Fermi energy can
approximately take the form
\begin{eqnarray}
\chi(\bm{q}=0;\epsilon_F;T) \propto
\log{|\mu_{-}/\Delta|}-1+(\mu_{+}/k_BT)\log{|\mu_{+}/\mu_{-}|}
~~~~~~~( for \hspace{3mm}\mu_{+} \ll \Delta). \label{4.32}
\end{eqnarray}
here $\mu_{\pm}=\epsilon_F\pm k_BT/2$. For finite $q$, we expect
that $\chi(q)$ deviates from $\log{q}$ in regime $q\leq
\sqrt{2m|\epsilon_F-k_BT/2|}$, which means the temperature has
notable affection on the diamagnetism of a neutral BLG when the
typical length scale exceeds $2\pi/\sqrt{2mk_BT}$, about $70\mu m$
at $T=1$K.

All of the above, we assume an ideal 2D BLG electron gas and
ignored the distance $d$ between up layer and bottom layer of
bilayer graphene. In order to obtain an analytic expression of the
susceptibility of bilayer graphene, we assume the wave number
$k_F$ and $q$ is much smaller than the cutoff wave number
$\Lambda$. An alternative farther investigation can be focused on
a more general case that $k_F$ is commensurate to or even larger
than $\Lambda$.

\section{Conclusions}

In this article, we study analytically the orbital magnetic
susceptibility of a weak-doped bilayer graphene (BLG) in spatially
smoothly varying magnetic fields by the low-energy Hamiltonian and
obtain the general analytic expression of the orbital
susceptibility of BLG, with finite wave number and Fermi energy,
at zero temperature. The induced magnetization, electric current
by the nonuniform magnetic fields in BLG are studied which are
different from that of MLG and 2DEG, but the argument, that the
magnetic object placed above the BLG is repelled by a diamagnetic
force which is equivalent to a force produced by mirror image on
the other side of BLG, still be approximately hold, only by
replacing the constant reduced amplitude with a reduced amplitude
dependent with the typical length of the system.
Logarithmically-dependent behaviors are found extensively exist in
both the orbital magnetism and the induced physical quantities by
specific external field. BLG shows crossover behaviors in the
responses to the external magnetic field as the intermediate
between MLG and 2DEG. The weak Fermi wave number dependent
behaviors, as a distinctive electric property of BLG, are found in
induced magnetization and electric current.

\section*{Acknowledgement}

This work is supported by NSFC Grant No.10675108.

\end{document}